\documentstyle[floats,aps]{revtex}

\begin{document}
\draft

\catcode`\@=11 \catcode`\@=12 \twocolumn %
[\hsize\textwidth\columnwidth\hsize\csname@twocolumnfalse\endcsname
\title
{A Novel Meron-induced Pseudospin Wave in Bilayer Quantum Hall
Coherent State and the Residual Zero-bias  Peak in Tunneling
Conductance}
\author{Yue Yu}
\address{Institute of Theoretical Physics, Chinese Academy of Sciences, P. O. Box 2735,
Beijing 100080, China}

\vspace{0.1in}

\maketitle
\begin{abstract}
In the bilayer quantum Hall coherent state for $\nu_T$ deviating
slightly from one, we show that, instead of the global order
parameter, the spontaneous breaking of the pseudospin $U(1)$
rotational symmetry is reflected by the periodic domain structure
accompanying with the charged meron pairs. The motion of meron
pairs induces a novel pseudospin wave. The long range order of the
periodic domains in a low bias voltage range leads to the residual
zero-bias peak in the tunneling conductance even when the
pseudopsin Goldstone feature in a high bias voltage range can be
distinct from it.

\end{abstract}

 ]

The low-lying gaped and gapless excitations in the various quantum
Hall systems played important roles in understanding the essential
physics of these systems ( For review, see \cite{q1,q2}). Among
them, the most remarkable one was the Laughlin quasiparticle,
which has a fractional charge and fractional statistics
\cite{Laugh}. The magnetoroton revealed the similarity between the
Laughlin liquid and the $^4$He superfluid \cite{gima}. The gapless
edge fluctuations distinguished the quantum Hall states from an
ordinary insulator \cite{halp} and showed a Luttinger liquid
behavior \cite{wen1,chang}. The particle-hole continuum of the
composite Fermi liquid caused the anomalous propagation of the
surface acoustic wave at $\nu=1/2$ \cite{hrl,will}. The skyrmion
spin texture exhibited a fruitful spin structure in the
multi-component quantum Hall systems \cite{son}. The
binding-unbinding of the meron-pairs in the bilayer quantum Hall
system gave the first example of the finite temperature phase
transition in the quantum Hall systems \cite{moon,murphy}.

Recently, a pseudospin collective mode \cite{wen} in the bilayer
spontaneous quantum Hall coherent state at $\nu_T=1$ has been
observed experimentally \cite{spie1,spie2}. This can be understood
as a pseudospin Goldstone mode due to the spontaneous breaking of
symmetry of particle number difference between two layers
\cite{Fer,wen}. Accompanying with this linear dispersing
collective mode, there existed a Josephson-like tunneling between
layers. These intriguing experiments have renewed the theoretical
research interest greatly
\cite{sch,stern,bal,stern1,fog,dem,kim,jog,vei,buk,iwa,yu}  while
opening a variety of unsolved issues ( for a short review, see
\cite{gir}).

Among these issues, the most urgent two are: First, differing from
the Josephson effect in a superconductor junction, the zero-bias
conductance peak in the experiment has a finite width and height
even at the zero-temperature. Second, in the common Josephson
effect, the position of the tunneling peak moves as a magnetic
field perpendicular to the tunneling current is applied. In the
bilayer quantum Hall case, this Goldstone feature did appear as a
parallel field is applied while the central peak remains
unexpectedly \cite{spie2}. In this Letter, we shall focus on the
second issue. We see that a new excitation induced by the motion
of meron pairs may cause this phenomenon.

Where does this new excitation come from? In realistic samples,
there are long-range density fluctuations with about the relative
magnitude of $ 4\%$ as mentioned in \cite{stern1}. This
corresponds to charged meron pair excitations such that the total
filling factor $\nu_T$ can deviate from one. When the tunneling is
turned on, beside the logarithmic interaction between paired
merons, there is a linear string confining energy. When the
tunneling is small enough, the logarithmic one dominates and
determines the optimal separation between two merons constituting
a pair. The linear term may distort the pesudospins.(Hereafter we
equal the word 'spin' to the pseudospin.)  In the picture of
\cite{moon}, only the spins near the string are distorted while
the spins far from the string array in the same direction. As we
shall see in this Letter, a more realistic spin configuration
induced by the meron pairs is a periodic varying spin
configuration. Since the $U(1)$ symmetry is spontaneously broken,
this periodic configuration will want to lie on a preferred
direction. This leads to a long range order of the periodic spin
configuration. When the meron pair moves, the spin configuration
travels simultaneously. If all charged meron pairs drift in a bias
voltage, the motion of the spin configuration forms a spin wave.

Why does this induced spin wave indicate a zero-bias peak in the
tunneling conductance? When the meron pairs drift, the induced
spin wave travels in a wave velocity $v_{isw} \leq v_m$, the meron
drift velocity which is proportional to the bias voltage $V$. This
spin wave contributes to the tunneling current response function a
pole $eV\pm eA|V|$ for a constant $A$, which means a zero-bias
peak in the conductance.

Why can the Goldstone mode be seen while this zero-bias peak still
exists? In very low bias, only the meron induced spin wave
contributes to the tunneling current response function. Since the
parallel field can only change the wave length $\lambda$ of the
induced spin wave but does not change the drift velocity of the
meron pairs, the zero-bias pole contributed by the induced spin
wave was not shifted. As the bias voltage increases, the meron
pairs are accelerated. If the meron pairs move so fast that the
spin configuration can not respond, the global order parameter is
restored because the spins which are not right beside the meron
pair can not see the pair. Thus, the Goldstone mode recovers in
the relative high bias, which contributes to the response
function. When a parallel field $B_\parallel$ is turned on, the
pole of the Goldstone mode shifts to $eV\propto B_\parallel$.
Namely, we have two energy scales, while the residual peak exists
in the low bias, the Goldstone feature is shown in the high bias.

The finite height and width will not be discussed in details.
Several scenarios of the disorder source have been provided
\cite{bal,stern1,fog} but a microscopic understanding is still
lack. In the present model, the induced spin wave may be
dissipative since it is much easier for the meron pair scattering
to influence the long range order of the periodic spin
configuration than a global order parameter. Thus, it is
anticipated that the central peak has a finite height and width.
The another possibility is the charging effect of the
pseudoskyrmion \cite{iwa}.

Why does the height of this residual peak reduce as the parallel
field increases? As we shall see, the domain structure of the
meron pairs is controlled by a modulus $k$. For $B_\parallel=0$,
$k_0>1$ implies a finite domain period length. As $B_\parallel$
increases, $k$ decreases and eventually, at a critical $\tilde
B_\parallel$, $k$ is down to 1. This implies the period length
tends to the infinity and the induced spin wave is suppressed.
Namely, the height of the central peak of the conductance goes
down to zero monotonically as $k\to 1$.

In the pseudospin language, the order parameter is a unit vector
$\vec m=(\cos\varphi,\sin\varphi,m_z)$. The Hamiltonian for
$\nu_T=1$ layer-balanced coherent states when a parallel field
exists is given by \cite{q2,moon}
\begin{eqnarray}
H=\int d^2r\biggl\{\frac{1}{2}\rho_s|\nabla\varphi|^2-\frac{t
}{2\pi
 l_B^2}\cos(\varphi-Qx)\biggr\}, \label{Hami}
\end{eqnarray}
where $\rho_s$ is the spin stuffness \cite{moon}; $l_B$ is the
magnetic length; and $Q=\frac{edB_\parallel}{\hbar c}$ with the
layer spacing $d$; the gauge is chosen as $\vec
A_\parallel=xB_\parallel\hat z$; $t$ is the tunneling amplitude.
We turn off $B_\parallel$ first. The cheapest energy charge-$\pm
e$ excitation is the meron pair with the opposite vorticities and
the same charge ($\pm e/2$) \cite{moon}. In the absence of the
tunneling, the pair is confined by the logarithmic attraction. It
is called a skyrmion \cite{son}. When the tunneling is turned on,
the skyrmion may be distorted and eventually turns into a meron
pair confined by a domain wall. The domain with an infinite length
string along the $y$ axis has its optimal form given by
$\varphi(\vec r)=2\arcsin[\tanh s]$, where $s=x/\xi$ with
$\xi=\sqrt{{2\pi l_B^2\rho_s}/{t}}$ \cite{moon}. In fact, the
domain structure may have a general form \cite{liang}
\begin{eqnarray}
\varphi(\vec r)=2\arcsin[k{\rm sn}(s,k)]+\varphi_0, \label{sn}
\end{eqnarray}
where $\varphi_0$ is a constant and ${\rm sn}$ is the Jacobian
{\it sine-amplitude} elliptic function. It is given by ${\rm
sn}(s,k)=k^{-1}{\rm sn}(ks,k^{-1})$ if the modulus $k>1$. The
period of ${\rm sn}(ks,k^{-1})$ is $4K(k^{-1})/k$  for $k>1$ where
$K(k) $ is the first kind complete elliptic integral. When $k\to
1$, ${\rm sn}(s,1)=\tanh s$, going back the solution in
\cite{moon}. The string tension may be calculated by \cite{chai}
\begin{eqnarray}
{\cal T}_0(k)=-\frac{\rho_s}{R}\int d^2\vec
r\varphi\nabla^2\varphi={\cal T}_0I(k),\label{st}
\end{eqnarray}
where $I(k)=\frac{k}{2}[2 E(k^{-1}) -\pi\sqrt{1-k^{-2}}]$; $E(k)$
is the second kind complete elliptic integral and ${\cal
T}_0=8\rho_s/\xi$. In the limit of $k\to 1$, $I(1)=1$ and ${\cal
T}_0(1)={\cal T}_0$. $I(k)$ monotonically decreases as $k$
increases. If $k\to \infty$, $I(k)$ and ${\cal T}_0(k)\to 0$. For
$k<1$, $E(k^{-1})$ is imaginary and (\ref{st}) has no physical
meaning and the only meaningful solution is the trivial one (k=0).
The optimal separation between merons in a pair is given by
$R_{s0}=e^2/8\pi\rho_s\epsilon$ for the skyrmion or by
$R_0(k)=\sqrt{e^2/4{\cal T}_0(k)}$ for the domain wall
\cite{moon}. If $R_0<R_{s0}$, the minimal pair energy is given by
\begin{eqnarray}
E^{min}_{\rm pair}(k)=\frac{e^2}{4\epsilon R_0(k)}+{\cal
T}_0(k)R_0(k)= \sqrt{e^2{\cal T}_0(k)/\epsilon}.\label{min}
\end{eqnarray}
For $R_0>R_{s0}$, the $k$-dependent minimal energy of a pair may
be approximated by
\begin{eqnarray}
E_{\rm skyr}^{min}(k)\approx 2\pi\rho_s(1+\ln R_{s0}/R_{\rm
mc})+{\cal T}_0(k)R_{s0},\label{shyr}
\end{eqnarray}
where $R_{\rm mc}\sim l_B $ is the meron core size. Although this
infinite wall result may not be quantitatively correct for
$\xi>R_{s0}$, it may grab the qualitatively $k$-dependent
behavior. In (\ref{min}) and (\ref{shyr}), the core energy
$2E_{\rm mc}$ has been omitted. And both of them seem to imply
that the state $k\to \infty$ is favorable. However, the domain
stores the $k$-dependent energy given by substituting (\ref{sn})
to (\ref{Hami})
\begin{eqnarray}
E_{\rm domain}(k)=\frac{At}{2\pi l_B^2}(2k^2-1). \label{dw}
\end{eqnarray}
In the realistic samples, the area $A$ that a meron pair occupied
is finite. Thus, one has to minimize
\begin{eqnarray}
E_{\rm total}(k)=E^{min}_{\rm skyr/pair}(k)+E_{\rm
domain}(k),\label{tot}
\end{eqnarray}
in $1\leq k< \infty$. The above discussion is valid if the spacing
between meron pairs is larger than the separation of two merons in
a pair because the interaction between pairs should be negligible.
 In the real samples, $\rho_s\sim 0.4$K and
$n_0\sim 5.0\times 10^{10}$cm$^{-2}$, a meron pair occupies an
area $\sim 72~l_B^2$; $t\sim 6.0\times 10^{-7}$(in the unit
$e^2/\epsilon l_B$). Using (\ref{shyr}) and (\ref{tot}), one has
$k_0=3.06$. The space period $\lambda=4\xi K(k_0^{-1})/k_0\sim
496l_B$, about 50 times of the meron pair spacing. To use
(\ref{min}), $t$ is restricted to $0.0016<t<0.1$. Taking, for
example, $t\sim 0.005$, one has $k_0\sim 1.04$, $R_0\sim 4.67 l_B$
and $\xi=2.65 l_B$. The space period $\lambda\sim 27.57 l_B$, 1.8
times of the domain length. In both large and small $t$ cases, we
see there are periodic spin configurations which destroys the
global order parameter. As we have mentioned, the spontaneous
breaking of U(1) symmetry may lead to all these domains extending
in the direction ( say $\hat e$) along which the spatial average
of the order parameter field has a maximal value. Furthermore, the
continuation of the order parameter field may require all domains
connecting smoothly by self-consistently adjusting the position
and $\varphi_0$ of each pair. This sets up a long range order of
the periodic spin configuration, e.g.,
$\langle\psi^\dagger_\uparrow(\vec r)\psi_\downarrow(\vec
r)\rangle\sim e^{i\varphi(r\hat e)}$ where $\varphi(r\hat e)$ is
periodic along the $\hat e$-direction except in the position of
the singular merons. However, the larger $t$ case will not be easy
to be observed because the number of domains in a period $\lambda$
is too small to self-consistently adjust the positions of the
meron pairs, which costs the Coulomb energy between the pairs.

Now, turn on the parallel field in the $y$-direction. Rewriting
the Hamiltonian (\ref{Hami}) by \cite{q2,moon}
\begin{eqnarray}
H=\int d^2r\biggl\{\frac{1}{2}\rho_s[(\partial_x
\theta+Q)^2+(\partial_y\theta)^2]-tn_0\cos\theta\biggr\},
\end{eqnarray}
where the changed variable $\theta=\varphi-Qx$. Since the extra
non-constant term is a total divergence, the domain structure of
$\theta$ is the same as (\ref{sn}) but the favorable lying
direction of the walls is fixed in the $y$-direction
\cite{q2,moon}. Thus, the Goldstone mode is suppressed.In a very
low bias, the periodic domain will move simultaneously as the
meron pairs drift. This induces a spin wave, which has the wave
velocity $v_{isw}\approx v_m$, wave length $\lambda$ and wave
vector $q_{isw}=h/\lambda$. Instead of the suppressed Goldstone
mode, this induced spin wave will contribute to the tunneling
current response function. As the bias voltage increases, the
response of the induced spin wave to the meron motion becomes
slow. And eventually, the spin wave can not respond to the merons.
Namely, the periodic domain structure disappears and the Goldstone
mode is recovered. We can call this the meron unscreening. The
unscreening voltage can be determined by the relaxation time
$\tau_1$ of the spin wave. However, this relaxation time is
different from the relaxation time $\tau_\varphi$ with a disorder
source. $\tau_1$ may be thought as a longitudinal relaxation time
while $\tau_\varphi< \tau_1$ is the transverse relaxation time
\cite{kittel}. The transverse $\tau_\varphi$ has been
microscopically calculated in \cite{jog} while there is no a
microscopic estimate to $\tau_1$ yet. However, in our case,
$\tau_1$ may be longer than $\tau_\varphi$ several orders because
$\tau_\varphi$ indicates the time of a perturbed local spin back
to the equilibrium state while $\tau_1$ is the response time of
the spin wave following the change of equilibrium state. In the
sample of Spielman et al used,
$\delta_\varphi=\hbar/\tau_\varphi\sim 0.75$K. If we assume
$\tau_1\sim 10^3\tau_\varphi$, the unscreening voltage $V^*$ can
be estimated by $v_m(V^*)\sim 10l_B/\tau_1$. We take the meron
pair density $n_m\sim 0.04n_0$ and $n_0=5.0\times 10^{10}$
cm$^{-2}$; the longitudinal resistivity $ \rho_{xx}\sim$
1k$\Omega$ and the sample linear size $\sim$ 1mm. Then, $V^*\sim
50\mu$V, coinciding with the bias voltage in which the Goldstone
feature appears in the experiment.

The suppression of the spin wave by a parallel field may be
understood as follows. For $B_\parallel\ne 0$, the string tension
decreases linearly, i.e., ${\cal T}(k)={\cal
T}_0(k)(1-B_\parallel/B^*_\parallel)$, where $B^*_\parallel$ is
the critical field of the commensurate-incommensurate phase
transition \cite{moon}. The optimal value of $k=k_0$ decreases as
${\cal T}$ goes down, and at $\tilde B_\parallel$, $k_0\to 1$. For
$t=6.0\times 10^{-7}$ but ${\cal T}=0.5{\cal T}_0$, $k_0=2.38$ and
$\lambda\sim 667 l_B$. $k_0=1$ arrives at ${\cal T}=0.001{\cal
T}_0$ and $\lambda\to \infty$. That is to say, at
$B_\parallel=\tilde B_\parallel\lesssim B^*_\parallel$, the
induced spin wave is destroyed and one has only the Goldstone mode
contributes to the response function. Then the central peak
disappears. For larger $t$, e.g., $t=0.005$ but ${\cal
T}=0.005{\cal T}_0$, one has $k_0\to 1$. Furthermore, for a
$B_\parallel>\tilde B_\parallel$, (\ref{min}) and (\ref{shyr})
become $k$-independent but (\ref{dw}) favors $k=0$. This gives
$\theta(\vec r)=0$, namely, $\varphi(\vec r)=Qx$, the commensurate
state. At $B_\parallel=B^*_\parallel$, the string tension vanishes
and a commensurate-incommensurate transition appears \cite{moon}.

Based on the above discussion, we now calculate the tunneling
current in two bias ranges. In the low bias range ($V<V^*$),
recalling the expansion $\arcsin[{\rm sn}(u)]=\frac{\pi
u}{2K}+2\sum_{n=1}^\infty
\frac{1}{n}\frac{g^n}{1+g^{2n}}\sin\frac{n\pi u}{K}$ with
$g=e^{-\frac{\pi K'}{K}}$, one can decomposite $\theta(\vec
r)=\theta_m(\vec r)+ q_{isw}x+\tilde\theta(\vec r)$ , where the
first term comes from the singular merons and the other two from
the domain. It implies that for an infinite wall, the
time-dependent $\tilde\theta(x\pm v_{isw}t)$ is the solution of
the equation of motion
$\frac{1}{v_{isw}^2}\partial_t^2\tilde\theta-\partial_x^2\tilde\theta=0
$.
 Hence, the effective Lagrangian of the realistic system for
$\omega=eV/\hbar$ reads
\begin{eqnarray}
{\cal
L}&=&\frac{\rho_s}{2}\biggl[\frac{1}{v_{isw}^2}(\partial_t\tilde\theta(\vec
r,t))^2- |\nabla\tilde\theta(\vec r,t)|^2\biggr]\nonumber\\
&&-\frac{t}{2\pi l^2_B}\cos(\theta_m(\vec r,t)+\tilde\theta(\vec
r,t)+q_{isw}x-\omega t).\label{la}
\end{eqnarray}
This effective theory is the same as the high bias one by the
correspondence $v\leftrightarrow v_{isw}$, $Q\leftrightarrow
q_{isw}$ and $\varphi\leftrightarrow \tilde\theta$ \cite{bal}. The
tunneling current now can be calculated in a similar manner in the
literature \cite{bal,stern1,fog}. To be specified, we take the
calculation result of the tunneling current in the version of
Balents and Radzihosky \cite{bal},
\begin{eqnarray}
J_Q(V)= \frac{N_0}{l_Bq_{isw}k_BT}\sum_s
s\biggl|\frac{k_BT}{eV-s\hbar v_{isw}q_{isw}}\biggr|^{1-\eta},
\label{j}
\end{eqnarray}
 where $N_0$ is a constant \cite{note1} and
$\eta=k_BT/2\pi\rho_s$ is the Kosterlitz-Thouless exponent. In a
very low bias, the velocity $v_{isw}\approx v_m\propto |V|$.
Eq.(\ref{j}), then, implies a zero-bias peak in the condcutance.
As $V$ increases, the response of the induced spin wave to the
meron pair motion becomes slow. And so the wave velocity reduces a
factor which is less than one, i.e., $v_{ism}<v_m$. Thus, the
current and the conductance reduce. For $V\to V^*$, $v_{isw}\to 0$
and the sum of $s$ in (\ref{j}) is zero and the current and
conductance vanish. The $Q$-dependence of (\ref{j}) is included in
$q_{isw}=h/\lambda$. For $B_\parallel=\tilde B_\parallel$, $k_0\to
1$ and $q_{isw}\to 0$. Hence, $J_Q(V)\sim O(q_{isw})\to 0$. This
indicates the zero-bias peak in the conductance is totally
suppressed when $B_\parallel=\tilde B_\parallel$.

If $V>V^*$, only the Goldstone mode contributes to the tunneling
current. The tunneling current in this bias range has been
calculated by several authors \cite{bal,stern1,fog}. In
\cite{bal}, the effective Lagrangian and the tunneling current are
the same as (\ref{la}) and (\ref{j}) by the correspondence
mentioned in the last paragraph. The phenomena for two bias ranges
are sketched in Figure 1 and resemble what were observed by
Spielman et al in their experiment \cite{spie1,spie2}.

In conclusions, we have found, in a low bias, a long range order
of the periodic domains and a meron-induced spin wave instead of a
global order parameter and the Goldstone mode, if $\nu_T$ slightly
deviates from 1. This leads to a residual conductance zero-bias
peak. In a high bias, this induced spin wave disappears as the
meron pairs are unscreened and the Goldstone feature is recovered.
In a critical parallel magnetic field, this spin wave can also be
suppressed. This explains the experimental results in
\cite{spie1,spie2}. We have assumed that the induced spin wave is
perfect. However, the real spin wave shape may be distorted
severely due to the merons and their scattering. This is one of
the reasons for the finite zero-bias peak.

The author would like to thank Z. B. Su for the thorough reading
of the manuscript and the useful discussion. He is benefit from
the valuable discussions with A. H. MacDonald, Q. Niu, Z. Q. Wang,
T. Xiang, X. C. Xie, L. You and F. C. Zhang. This work was
supported in part by the NSF of China.

{\centerline {Figure Caption}}

{\noindent Fig. 1 Schematic tunneling conductance for $T=0$ in two
bias voltage ranges.The tunneling conductance in $V<V^*$ is given
by the derivative of (\ref{j}). $\delta_1$ and $\delta_\varphi$
($\delta_1\ll\delta_\varphi$) have been added in the denominators
to round the peak. The Goldstone feature (or the derivative shape)
in $V>V^*$ is from the correspondence of (\ref{j}) \cite{bal}.}

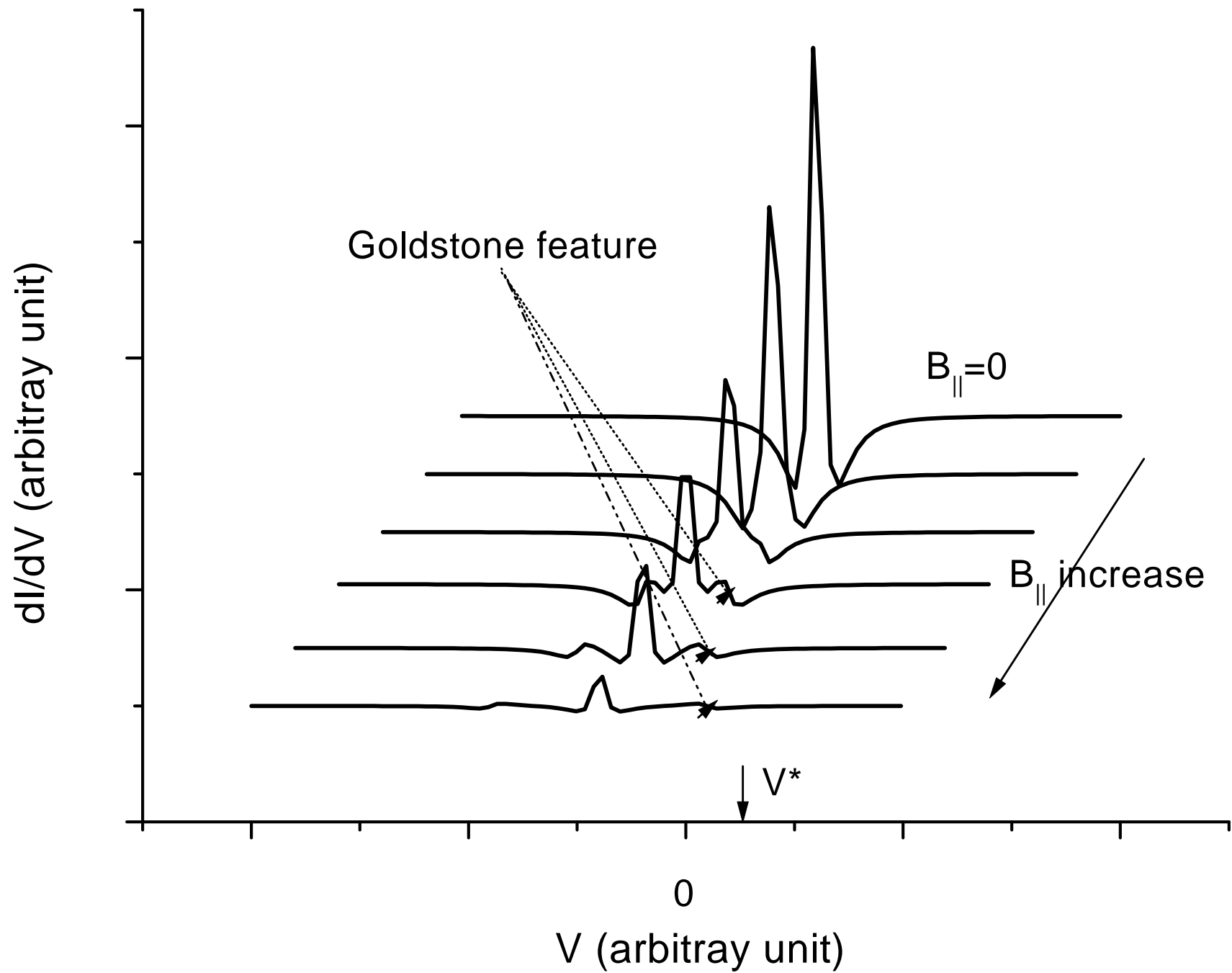
\end{document}